# Physiological complexity of EEG as a proxy for dementia risk prediction: a review and preliminary cross-section analysis

Milena Čukić Radenković[1], Simon Annaheim[1], Patrick Eggenberger[1,2,3], and René Michel Rossi[1,2]

[1] Empa Swiss Federal Labs for Materials Science and Technology, Laboratory for Biomimetic Membranes and Textiles, St. Gallen, Switzerland
[2] Institute of Human Movement Sciences and Sport, Department of Health Sciences and Technology, ETH Zurich Zürich, Switzerland
[3] OST, Eastern Switzerland University of Applied Sciences, Department of Health, Institute of Physiotherapy, St.Gallen, Switzerland

Mild cognitive impairement (MCI) is recognized as a predementia stage. It is an important risk factor for Alzhemer's disease (AD). AD is considered to be a connectivity disease due to the massive loss of connections between neural cells. Electroencephalogram (EEG) is the oldest brain imaging technique, is well documented, noninvasive and accessible in every modern hospital. The EEG-signal reflects activity of the brain cortex cells, which is very complex, as well as brain structure and organization. Contrary to the classical (standard) Fourier-based approach to analysis that presumes the signals' stationarity, fractal and nonlinear approach to analysis is much better suited for EEG analysis allowing us to early detect changes in such a complex dynamical system as the brain. The core of application of complex systems dynamics theory in physiology (physiological complexity) is connected to the stereotypy of disease that clinicians use to classify patients demonstrating similar clinical signs. Knowing that a healthy organism is expressing immense variability in its physiological parameters' range (which is a sign of healthy adaptability), it is well justified that the reduction in this physiological complexity is correlated with the disease (decomplexification). Certain levels of decomplexification are confirmed in healthy human ageing, but the levels of decomplexification characteristic of a disease are distinct and can serve as a marker. In early detection of dementia risk (and AD in particular), nonlinear measures extracted from EEG as a proxy of the brain as a complex system are promising as accesible, accurate and potentially clinically useful biomarkers of dementia. The aim of this work is to give the readers a review (perspective) of prior work on this kind of complexity-based detection from resting-state EEG and present our preliminary cross-section analysis results on how EEG complexity of supposedly healthy senior persons can serve as an early warning to clinicians. Together with the use of wearables for health, this approach to early detection can be done out of clinical setting improving the chances of increasing the quality of life in seniors.

**Key words**: Dementia, Alzheimer's, Mild congitive impairment, Physiological complexity, Early detection

## Introduction

One of the most terrifying things to happen for any ageing individual is dementia and Alzheimer's disease in particular, covering between 50% and 70% of all dementias, (Rossini et al., 2022). The risks are rising, as is rising of the number of older citizens. Characteristics of Alzheimer's disease (AD) are loss of neural cells, brain networks disconnection, neurofibrillary tangles, beta-amyloid plaques, synaptic traffic failure, and synaptic pruning, which are forming for years before the onset of clinical signs. The evidence from structural magnetic resonance imaging (sMRI) suggests that cognitive control while aging is associated with altered activation within the frontal, parietal, and occipital lobes (Xia et al,. 2022) (occipital are most likely to change first). Once clinical signs are there, it is too late to intervene due to irreversible damage to neural tissue. The problem is, that there is not a single developed medication or a specific medical procedure that can halt the development of AD, but early detection can offer precious time for both doctors and patients to apply some of the strategies (cognitive training, exergames, dancing etc) that are known to slow down the progression of the disease. The time window for hope is to find a way to act during the preclinical/prodromal phase where plastic brain reorganization and slowing down of connectivity issues and neural loss are still possible (Rossini et al., 2022).

Mild cognitive impairment (MCI) is often recognized as a syndrome, where the aging person is functional in regard to everyday life tasks, but her/his cognitive performance is deteriorating (Flicker et al., 1991; Petersen et al., 1995, 2001; Gao et al., 1998). There are two subgroups, amnestic MCI (aMCI) and non-amnestic MCI (naMCI), based on whether or not memory is impaired (Petersen et al., 2014). Amnestic MCI (aMCI) is considered to be a prodromal phase for AD (Rossini et al., 2022; Xia et al., 2023), the other one is correlated to different forms of dementias, like vascular dementia. Half of the people who exhibit aMCI are converted to AD (on average 3 years after the first MCI identification), so aMCI is considered to be a huge risk for AD and sometimes is called MCI-prodromal-to-dementia (Rosinni, 2022). Among other things we know are related to this prodromal phase, are high plasma concentrations of non-ceruloplasmin cuprum; free cuprum is demonstrated to affect cuproptosis (cell death), disruption of maturation of synapses, energy consumption in neurons and other important basic physiological mechanisms that are confrimed to be disrupted in AD (Squitti et al., 2006, 2013, 2016). Those high concentrations of free copper in serum fortunately could be reversed (by use of specific zinc-based formulations that can capture free copper) once detected (Squiti, 2016) and potentially increase the quality of life of a patient. Squitti and colleagues (2016) showed that high concetrations of free coper in plasma (>16μM/L) triple the estimated risk of AD. Other standard AD tests, including biochemical analyses from blood and cerebrospinal fluid, in combination with various cognitive tests such as the Mini Mental State Examination – MMSE (Ihl et al, 1989) are currently applied. Since a plethora of research of potential significance of various biochemical or cognitive biomarkers exists, and none of them are actually standard in clinical setting, it is beyond this format to discuss about limitations of those approaches; in general, we should be aware that each clinical and research institution might have specific sets of tests for establishing the AD diagnosis.

Many modern imaging techniques are able to detect known neuroanatomical/structural changes associated with AD (structural magnetic resonance imaging - sMRI, positron emission tomography - PET, magnetoencephalograpy -MEG, etc.), but the majority of those exams are extremely expensive, thus not available to everyone even in most developed western countries. For sorely needed early MCI detection that could be used for screening of all older citizens, the approach should be cheap, accurate, and accessible. A well-documented (and the oldest) neuroimaging technique that is relatively cheap, noninvasive, and accessible in almost any modern hospital (Niedermayer, 2004), the electroencephalogram (EEG), might fit this description of early detection method.

The aim of this work is to provide a review (perspective) of prior work on complexity-based detection from resting-state EEG and present our preliminary cross-section analysis results on how EEG complexity of supposedly healthy senior persons can serve as an early warning to clinicians.

**Characteristics of EEG changes in AD and MCI**

There is a body of knowledge on a number of confirmed changes in EEG in AD. Characteristic structural neuropathology that involves neural cells loss, neurofibrilar tangles, amyloid plaques and metallomics markers, and aging-related changes in functional segregation and integration are mirrored in EEG features of a person. In regard to EEG sub-bands, AD patients exhibit the so-called 'slowing' of brain activity, which is visible as an increase in EEG power in low-frequency rhythms (delta, theta, and low alpha band). Also a significant decrease in EEG power and cortico-cortical coherence in higher frequency bands (high alpha and beta) was confirmed (Jeong et al., 2004). Fraction of serum copper unbound to ceruloplasmin (free copper) is also positively correlated with temporal and frontal delta EEG sources, regardless of the effects of age (Babilloni et al, 2007). The problematic part of those specific EEG features is that those changes are present in a lesser extent in healthy aging people (Smits et al., 2016). Specifically, the comparison of the levels of exhibited EEG complexity of young controls (YC), healthy aging people (HAP), MCI, and AD shows that these are possibly different phases of temporal evolution of the aging brain (Sun et al, 2020). It is demonstrated that from young age to maturation the complexity of EEG rises to a certain plateau and then falls slowly in healthy ageing (after mid-sixties), to become significantly lower in MCI and even extremely (pathologically) low in AD (Smits et al., 2016).
There is plenty of evidence on EEG differences between AD and healthy controls (HC) (Maurer and Dierks, 1992; Szelies et al., 1992; Leuchter et al., 1993; Schreiter-Gasser et al., 1993). Some of the features of AD EEG are an excessive delta (0–4 Hz) and a significant decrement of posterior alpha rhythms (8–12 Hz; Dierks et al., 1993, 2000; Huang et al., 2000; Ponomareva et al., 2003; Jeong, 2004; Babiloni et al., 2004a). Those observed changes in AD EEG are confirmed to be related to aberrated regional cerebral blood flow (rCBF) and are also correlated to the scores typically assessed by MMSE. In the case of MCI (compared to HC), similar changes were confirmed, namely, an increase of theta (4–7 Hz) and a decrease of alpha power (Zappoli et al., 1995; Jelic et al., 1996; Huang et al., 2000). It is suggested that those changes are caused by the loss of cholinergic basal forebrain neurons projecting to the hippocampus and fronto-parietal areas (Holschneider et al., 1999; Mesulam et al., 2004). In addition, it is shown that the increase in delta rhythms characteristic of AD is most likely a consequence of atrophy in mesial-temporal posterior and/or frontal cortical areas (Fernandez et al., 2003; Babiloni et al., 2006a).

**Physiological Complexity and EEG**

It is well known that the human brain is among the most complex systems in nature. It is logical that besides the structural complexity, the complex dynamics of its millions of constitutive neurons and their numerous functional networks are also exhibiting very complex dynamics. The signal that is reflecting this complex dynamics is EEG, and it is a composite of contributions of many different and physically distant sources. EEG is a nonstationary, nonlinear, and noisy signal, that is seldom predictable; as a matter of fact, if it is regular/predictable it implies that something is wrong with the brain dynamics (like epileptic seizure, for example). Networks of the brain are highly specialized and highly integrated (Tonnoni 1998). EEG signals actually represent interactions between varying elements of this highly complex system (Stam 2005). As with all other physiological signals (consequences of deeper interactions), EEG exhibits great variability and its fluctuations change in a characteristic way; a certain pattern in brain activity can be recognized, and it shows self-similar properties (Stam 2005).
Fractals are, put simply, non-euclidean geometrical forms (opposed to familiar Euclidean integer dimensions, fractals have non-integer dimensions - for example, 1.67). Think about the shape of a leaf, or cauliflower, or the shape of tree branches, the pattern of a river, the shape of a coast, a cloud, a mountaintop, as well as the shape of Purkinje cells. None of the mentioned natural forms can be described by a straight line, or a plane, or by any standard euclidean figures; in fact, all natural forms are fractals, whereas Euclidean geometry is just a simplified description of our reality. As Benoit Mandelbrot (who coined the term fractal - from latin fractus/broken, in 1967) noticed 'the lightning is not traveling in the straight line, nor is the dog bark, the mountains are not cones, and the clouds are not spheres...'. Among those natural forms we can observe that the pattern that characterize them are the same on all scales. If you magnify the rugged curve that illustrates, for example, a graph of electric signal recorded from your heart- (electrocardiogram, ECG) on different magnifications those shapes look alike. One piece of a signal is a reminisence of another. When we try to quantify it with mathematical tools we can say that those fractals are exact or statistically similar. The former is characteristic for generated mathematical graphs, but the latter is true for any natural shape or form. These mathematical descriptions originated from the early work of mathematicians from 18th and 19th century (Sirepinski, Koch, Cantor, Weierstrass, Lucia, Peano etc.) and they were called degenerate functions, even 'pathological' functions since it is not possible to analyse them as standard mathematical functions. Fractal dimension can also be a measure of how a geometrical structure is filling the space, and all physiological structures are actually maximizing space use (think of how human brain folds and gyri fit the limited volume of the skull). When we calculate fractal dimension of a signal, we are estimating how complex it is in time domain, how 'wrinkled' or 'rough' it is. The most appropriate algorithm for calculating fractal dimension in physiology is Higuchi fractal dimension - HFD (Higuchi, 1988) and it is basically estimating the complexity in time domain. A lot of work is done on electrophysiological signals and its use in medical applications.

Self-similarity is a characteristic fractal feature of human physiological forms and signals, and those could be quantified by fractal analysis (Stam 2005; Klonowski 2006, Buszaki, 2014). Namely, the neural activity of the brain is exhibiting similar features over and over again on a scale-free basis (Smits et al. 2016). Many authors showed that this complexity has a physiological purpose since it makes the system capable of adaptation to ever-changing internal and external conditions; a certain margin is always present and certain activities are a combination of a random variation (previously thought to be the noise, see Stam et al., 2005) and self-similar irregular processes (Goledberger et al., 2000). Once the system starts losing complexity it can be an early sign of aging and disease, Goldberger explained this with old medical reliance on stereotypy. All medical doctors use similar features to classify their patients as sufferers of certain diseases (Goldberger 1997). Pincus, Goldberger, Peng, Hausdorf, Klonowski, Stam, Eke and others demonstrated that it also applies on the field of physiological complexity. We call it decomplexification - a pathological loss of complexity (Goldberger 1997, Klonowski 1999).

The same corollary has already been used for EEG-based detection in several other disorders, like Parkinson's disease (Risannen, 2008), or depression (Ahmadlou 2012; Hosseinifard et al., 2014; Bachmann et al. 2018; Čukić et al., 2018, 2020, 2021); a possible explanation is that once a deep structure in the brain is damaged/compromised, the task that requires its involvement could be modified since the brain is compensating (compensatory mechanisms) for the loss by utilizing other available structures. And that is something we can measure from EEG signals. The level of complexity was suggested as a potential biomarker for early diagnosis of AD (Alturi et al, 2013; Smits et al, 2016; Polanco et al, 2018).

In the brain, the loss of complexity implies that the neural systems are becoming less flexible and less efficient in their collaboration and processing (Higuchi 1988, Zapasodi et al, 2014; Ahmadlou et al, 2011) hence, this could be used to detect AD.

**Fractal and nonlinear measures for MCI detection**

It is demonstrated that AD patients show a decline in complexity of neural activity, proven as reductions in fractal dimension (and other complexity measures of EEG) over the whole brain (Staudinger 2011, McBride 2014; Ahmadlou 2011). A cognitive impairment shows a further decline in complexity, as the disorder progresses (Jeong 2004; Mizuno 2010). It is interesting to note that healthy aging individuals also show a certain decrease of complexity in brain activity and in synchronization among the oscillating rhythms produced by different neural areas, as shown by several measures of complexity (Goldberger 2002; Gaal 2010).
Smits and colleagues (2016) examined how this decomplexification is developing from young healthy controls to healthy aging individuals to AD patients. First, they established the range of numerical values (the whole head EEG-extracted HFD) for certain homologous areas to check also for symmetry, since we know that, besides complexity loss, a characteristic asymmetry of EEG signals can develop (Figure 1 in Smits et al., 2016. Regional HFD and HFD Homologous areas symmetry (HArS) estimates). They found that the fractal dimension of EEG depends on age and is further reduced in AD. They also found that HFD increases from adolescence to adulthood and decreases from adulthood to old age (in the form of an inverted U-shaped curve) (Smits et al., 2016). They concluded that this decomplexification is a feature of the aging brain, and it worsens in AD, illustrating irreversible damage of underlying structure. In addition, they observed a regional prevalence of fractal dimension changes correlated to aging and AD. The effect was most prominent in parietal and central brain regions in healthy people (who age), and temporal and occipital complexity decline most strongly correlated to impairment in AD. Also, EEG-derived fractal dimension is declining with an increase of non-ceruloplasmin cuprum in AD patients (similar finding with Babiloni, 2007; Smits 2016). The loss of symmetry between homologous areas (parietal HFD) was depending on age (Smits 2016).
Other researchers that used all applied various entropy measures (one of the most prominent families of nonlinear measures in use in physiology), compared healthy controls, MCI, and AD, and found that indeed, MCI acts as a transition from normal aging to AD (Sun et al,. 2020). Entropy showed to be one of the most potent concepts to test the irregularity in time series; entropy-based methods combine the complexity of the signal with its unpredictability. Unpredictable signals are more complex than regular ones, because they are more irregular. Regarding above description of physiological complexity, healthy complex system should exihibit irregularity as it is linked to described innate adaptability to internal and external changes in important variables (Shannon, 1948). Van Neuman (1948) was using similar concepts as inititial description of Information theory to increse our understanding of biological systems dynamics. In 1948 he described some of the prevalent neurological and psychiatric disorders as diseases of loss of connections between previously functional networks and centers that must collaborate on a task. The alterations in how AD brain networks behave could be examined this way, too. The majority of successfull studies that used EEG, opted for using the resting state EEG (most frequently closed eyes) since these recordings provide more reliable estimates of brain adaptability (Kumar 2015; Lehman et al., 2014) and they reflect the contribution of networks with the highest metabolic acitivity (Ganzzeti, 2013). Abasolo (2005 and 2008) showed that, in all EEG channels, complexity of EEG in AD is below the one calclulated in MCI. Certain entropy measures clearly demonstrated decomplexification in MCI and AD; Approximate entropy (ApEn) (Abasolo 2008) and Sample entropy (SampEn) (Abasolo 2006, Woon 2007) in particular. A significant reduction of complexity at electrodes P3, P4, O1 and O2 (parietal, occipital and temporal regions) were consistently detected in AD when compared with healthy controls. When different entropy algorithms on EEG recordings were compared (Sun et al., 2020), AD and MCI patients had lower entropy values in the five regions (parietal, occipital, temporal, frontal, central) (AD<MCI<HC). This suggests that AD and MCI patients' EEG exhibit significantly lower complexity in the frontal, temporal and central regions compared to HCs. One of the most striking interpretations in entropy measures application to detect AD was that AD patients exhibit the lowest compexity and the greatest regularity, enabling the identification of this condition (Sun et al., 2020). Also, the fact that those complexity changes are detectable from all EEG electrodes suggest that the practical EEG-based early detection does not require all the electrode recordings (standard EEG postions), but a smaller number at the most important positions, like those from parietal, occipital and temporal regions.

Since one of the hallmarks of AD is that neural networks are disconnected due to synapses failing and cell loss, it was stated in the literature that AD could be understood as a 'network disease' or even 'disconnection' disease. Among other publications, Czoch et al. examined a particularly developed method of how one can explore brain changes in AD using network theory (Czoch et al 2023). They explored alterations in both functional brain connectivity and in the fractal scaling of neuronal dynamics, by applying a new concept of fractal connectivity (FrC) (to capture long-term interactions between brain regions). The authors introduced a novel multiple-resampling cross-spectral analysis (MRCSA), which allows for unbiased fractal characterization of not only regional neural activity but also fractal characterization of connectivity networks (Czoch et al., 2023). This analytic approach is promising since many of the discriminative neural patterns showed a strong association with cognitive functions that were found diminished in the elderly population.

The results of cognitive neuroscientific research illustrate what physiological complexity is telling us. Simple cognitive tests measuring, for example, the speed and accuracy when people are asked to solve a problem reveals that older people are not losing accuracy (compared to younger people), but they are slightly slower on the task. The interpretation (from neuroimaging) is that older people are simply using more resources, the same task requires several different neural networks to solve the problem which is again, a form of compensation. People who are experiencing MCI are most often aware of it, they know that their performances are dropping, but again, if they put more effort and attention to it, they can still do the task effectively although more slowly. Here comes the important role of strategies to keep MCI brain effective for longer: certain cognitive training, or better multimodal approach of exergames (Eggenberger et al. 2020), or dancing are requiring that several different systems stay activated for longer when the so-called neural-reserve is still there to be remodelled and recruited for the specific task (Rossini et al. 2022). The problem with AD is that compensation (or performing the same task with different program, or different neural pathways) is no longer possible, due to neurodegeneration (Sun et al., 2020). Some structures are unfortunately irreversibly lost. That boils down to the common sense knowledge that we are healthy until we can learn new things. Contrary to a priori expectations, our hippocampi are producing new cells until the end, if they are functional - that is to say, they are not affected by neurodegenerative processes characteristic for dementia. In addition, those cells that are newly built every day are staying functional only if you put them to use.

**EEG-extracted features and machine learning for detection**

A number of papers from the last couple years explored not only EEG-based detection but also a multimodal approach with other known biochemical and cognitive markers while employing machine learning. We are mentioning here just some of them, and will compare the results later in the discussion. Sharma (et al., 2019) used standard EEG measures (power spectral density, skewness, kurtosis, spectral skewness, spectral kurtosis, spectral crest factor) and spectral entropy (SpEn) and Higuchi fractal dimension (HFD) as features for Support Vector Machine (SVM) that seems to be the favorite machine learning model in this kind of research. In addition, they performed finger tapping tests (FTT) and continuous performance tests (CPT) (to measure motor speed and sustained attention) as tests that are not bound to educational barriers. Once more they confirmed that in dementia EEG is 'slowing', the complexity of EEG is reduced, and the synchronization of EEG signals is perturbed (Sharma et al. 2019). The authors report that for all the features their classifications reached high accuracy (over 73%) but in the end, the highest accuracy in the classification of dementia was for FTT (84% accuracy) and CPT (88% accuracy) which helped differentiate dementia from MCI. It is interesting to note that the features that led to higher accuracy of classification- CPT and FTT- are not EEG-extracted features. An interesting study by Engedal et al. (2020) aimed to examine whether quantitative EEG (qEEG) using the statistical pattern recognition (SPR) method could predict conversion to dementia in patients with subjective cognitive decline (SCD) and MCI. The sample comprised patients from five Nordic memory clinics.  They concluded that applying qEEG using the automated SPR method (accuracy 69%) could be helpful in identifying patients with SCD and MCI who had a high risk of converting to dementia over a 5-year period (Engedal et al., 2020). Movahed et al. (2022) used several machine learning models and several different sets of features (spectral and functional connectivity, plus nonlinear EEG measures) to detect MCI. The best performance was by linear SVM (accuracy 99.4%,) in recognizing MCI cases. Rossini and colleagues (2022) described the part of the AI-MIND project method, that uses Graph analysis tools - Small World (SW), combined with machine learning methods (SVM), to identify the distinctive features of physiological/pathological brain aging focusing on functional connectivity networks evaluated on electroencephalographic data and neuropsychological, imaging, genetic, metabolic, cerebrospinal fluid, blood biomarkers. To our best knowledge, the results of that large multicenter study is still not published; would be interesting to know how this multimodal approach translates to real clinical situations. Another interesting research paper from the same year, by Geng et al. (2022) proposed an EEG-based method for MCI detection, but the EEG was recorded during sleep. The authors were extracting specific features of sleep (slow waves and spindles features) to characterize neuroregulatory deficit emergent with MCI. They also used the SVM classifier and gated recurrent unit (GRU) network to identify MCI. In the end, the MCI classification accuracy of the GRU network based on features extracted from sleep EEG was the highest, reaching 93.46%. Experimental results show that compared with the awake EEG, sleep EEG can provide more useful information to distinguish between MCI and HC. The researchers showed that this method can not only improve the classification but also facilitate an early intervention for AD (Geng et al., 2022). In another recent study, researchers performed a two-step three-level classification of HC, MCI, and AD groups using linear discriminant analysis (LDA) and SVM (Jiao et al., 2023). As features, they used several EEG extracted measures standard spectral, Hjorth metrics (activity, mobility, and complexity), time-frequency property (STFT), sample entropy, and microstate property. In this work, predictions of MMSE tests were obtained using EEG feature only, and Cerebrospinal Fluid (CSF)/ apolipoprotein E (APOE) biomarkers, hybrid features (EEG, CSF/APOE, sex, and age) as the predictors of the regression model, respectively (Jiao et al., 2023). The authors reported EEG biomarkers as features achieved over 70% accuracy in the three-level classification. Meghdadi (et al., 2021) calculated power spectral density for each EEG channel and spectral coherence between pairs of channels, and compared AD group with age-matched controls. They found increase of both at the lower frequencies (delta, theta). A (small) significant increase was detected in MCI group at temporal regions. The authors defined Power Distribution Distance Measure (PDDM) as a distance measure between probability distribution functions to be used for automated predictions of conversion of MCI to AD.

In contrast to those publications that mainly used spectral measures of EEG for further classification, there are a number of papers that performed the task by using nonlinear measures as features. We also showed that once the EEG signal was characterized with fractal and nonlinear measures, practically any of the most popular classifiers yielded high accuracy in detection (Čukić et al., 2020).

**Table1**. Tabular summary of papers that used nonlinear measures for feature selection from EEG (in combination with ML models) to classify HC, MCI and HC with reported Sensitivity, Specificity and Accuracy.

| Method/Feature extraction | Sensitivity | Specificity | Accuracy | Publication |
|---|---|---|---|---|
| Visibility graph complexity (VG) | NR | NR | 97.7% | Ahmadlou et. al (2010) |
| ApEn, SampEn | NR | NR | 91.70% | Acharya et al (2012) |
| TsEn<br>LZC | 85,7%<br>100% | 84,6%<br>92,3% | 85%<br>95% | Al-Quazzaz et al. (2018) |
| Lempel Ziv (LZ)<br>Multiscale LZ | 80%<br>86,8% | 78,1%<br>84,3% | 78,5%<br>85,7% | Liu et al.(2016) |
| HFD | 66,87% | 100% | 80% | Al Naumi et al. (2018) |
| Multiscale entropy (MSE) | 88,71% | 69,09% | 79,49% | Fan et al. (2018) |
| EpEn | 88,7% | 100% | 91,6% | Houmani et al.(2018) |
| ApEn<br>SampEn | 90,91% | 63,64% | 77,72% | Simons et. Al (2018) |
| MSE | NR | NR | 73% | Chai et al. (2019) |
| SpEn, HFD | 86% | 81% | 84,1% | Sharma et al. (2019) |

All of them compared those three classes, healthy controls (HC), MCI, and AD, but for the features used Lempel Ziv Complexity (LZC), Spectral entropy(SpecEn), Fractal dimension (FD), Multiscale entropy(MSE), Approximate entropy (AppEn), Epoch based entropy(EpEn), Sample entropy (SampEn), Fuzzy entropy (FuzzEn), Higuchi fractal dimension (HFD), Tsalis entropy (TsEn), and multiscale LPZ algorithm (Chai et al 2019; Fan et al 2018; Houmani et al, 2018; Simons et al., 2018; Al-Nuaimi et al., 2018; Al-Quazzaz et al., 2016; Liu et al, 2016; Ahmadlou et al, 2011). As expected, all those studies that characterized EEG signals with nonlinear measures yielded high accuracies (from 79.49% to 97%). Acharya et al. were among the first teams that detected AD by use of SVM, with features APEn and SampEn, and reported overall accuracy of 91.7%. Ahmadlou et al. (2011) tested two different fractal dimension methods, Katz and Higuchi methods, for diagnosis of AD using EEG signals. They reported a high accuracy of 99.3% discriminating patients with AD and normal subjects using the Katz method. Sharma (2019) is the only one that used both spectral-based and nonlinear features, but it is difficult to estimate the method since they did not even report the numerical values of HFD, just test statistics of comparisons. It is very challenging to compare those very diverse combinations of characterization of signals and various machine learning models applied, plus their different samples and sample sizers. We cannot answer straightforward whether the accuracy of traditional features based classification were higher than those nonlinear-feature-based classifications. It is essential to think of how representational they are; or, how appropriate they are for characterization of this particular signal. We can speak here about the methodological approaches that address the task differently and suggest what would improve its practical utility. For example, if classification is performed with very high accuracy (> 90%) and the training set was very small, we can be pretty sure that on unseen data that accuracy would be much lower, or not informative at all. Some authors also used the mix of traditional and nonlinear measures to characterize the data, which makes this comparison even more difficult. The bottom line is that we should probably think about a kind of standardization when any ML is used on electrophysiological signals (for clinical application) to prevent overfitting and make mandatory internal and external validation, or cross-validation, and insist on much larger datasets as requirement for publication (Kohavi et al 1995; Ng, 1997; Efron&Tibshirani, 2020). What is even more recommendable is to rely on receiver operation characteristic (ROC)/area under the curve use, since in our opinion it is much more informative measure of model practical usefulness than accuracy.

**Our preliminary results on EEG-extracted HFD**

Based on the method that Smits (et al 2016) used to detect AD with HFD (although Babilloni and Ahmadlou performed similar studies in 2007 and 2011), we decided to perform a similar analysis and reuse previously collected data at our institute. The project was intended to examine what the best strategies are to help aging citizens stay in good cognitive and physical form ) Eggenberger et al. 2021. We used baseline EEG recording, in order to check the state least affected by the intervention. The aim is to replicate a part of the results of Smits et al. (2016) and examine how this fractal analysis approach can help us characterize a healthy aging population that did not have a diagnosis of cognitive impairment and could be screened out of a clinical setting. Since that was a multimodal data collection (ECG recording, EEG recording, multiple cognitive and motor tests, etc) on a group of seniors (n = 84) that are considered healthy, we opted for EEG analysis only for screening purposes. The details of the method could be found in Eggenberger et al., 2021.

In this kind of complexity analysis (fractal analysis) where specific algorithm for fractal dimension is used (Higuchi, 1988), a very important thing is to determine parameter $k_{max}$, which is affecting the curve reconstruction from EEG-extracted time series. In our prior research we used lower values of this parameter according to the finding what in human physiology the most optimal values (Spasic et al., 2005). In this work we intentionally followed the method described in Smits et al. (2016) to be able to compare our results with what they reported. We chose the same $k_{max}$ =65 as their justification was very well based on how they preprocessed the signal and decided on the number of sequences (see also Klonowski, 2007 and Kalauzi, 2012). We found lower EEG complexity levels in some participants than what is reported in Smits' research as AD levels. We know from previous experience that researchers seldom report all the details of their methodology, hence it is often challenging to compare the results (which is the chronic problem of reproducibility in science). Here we can say that we can confidently build on Smits and her colleagues' findings, since the methodology can be reproduced, and used for further research on a larger sample in search for the most accurate early screening for aMCI as prodromal phase of AD. The HFD results calculated from participants EEG suggested that some seniors that were considered to be 'healthy' at the time of data collection, have been detected to be at AD risk. Based on the previous research (Smits et al. 2016), healthy seniors present HFD values around 1.91 (the differences are sometimes on the third or fourth decimal), as calculated by exactly the same method, while younger healthy people exhibit higher HFD values (above 1.92 or 1.93). Those whose resting-state EEG is of complexity below 1.88/1.89 are (according to the literature findings) already exhibiting MCI in the initial phase, while those who have even lower values were already classified as AD (in previous literature applying HFD for detection). In this Dataset, those who have values lower than 1.85 for HFD, were compared with HC as a potential group at risk. For example, two participants showed to have abnormal complexity levels - decomplexification of their EEG time series on the majority of electrodes.

When we compared the healthy group to the suspected MCI, the significant statistical differences were confirmed in the following EEG positions: P7 ($p<0.001$), P3 ($p<0.001$), P4 ($p<0.002$), P8 ($p<0.001$), O1 ($p<0.001$), O2 ($p<0.001$), and Fp2 ($p<0.018$). According to effect sizes (ES) calculated for the series that demonstrated statistically significant differences, the most prominent positions from EEG recordings are P7, P8, P4, O1, and O2, and with just mild ES, P3, and Fp2. Cohen's d (equivalent to Hedge's g, with Bessel's correction) showed effect sizes that are considered very large (for P7 Cohen's d was 2, for P4 1.35, for P8 was 1.39) and large (for O1 0.689) while other prominent effect sizes were in the range of small effect sizes (P3 0.18, O2 0.225, and Fp2 0.183). According to Cohen (Cohen, 1988) 0.2 ES is considered small, 0.5 is considered medium, and everything above 0.8 was considered large effect size.

After test statistics and effect sizes examination, we applied Principal Component Analysis (PCA) that we use in prior machine learning projects to check for data separability (and the reduction of dimensionality). We relied on the first three PCA components as it is shown that they explain 75% of variability(Jolliffe, 2002). PCA showed that the data are clearly separable. As in prior research, decomplexification characteristic for AD was contrasted with healthy complexity exhibiting much more individual variability, important for adaptation to internal and external conditions of human physiology. Our aim here was to show that the data in this datased are *separable*, hence methodology could lead to plausible classification, employing an ML model for application (a number of models would sufice). From theoretical point of view LASSO (an algorythm for linear regression that uses shrinkage), embeded regularization or regularization networks would be ideal.

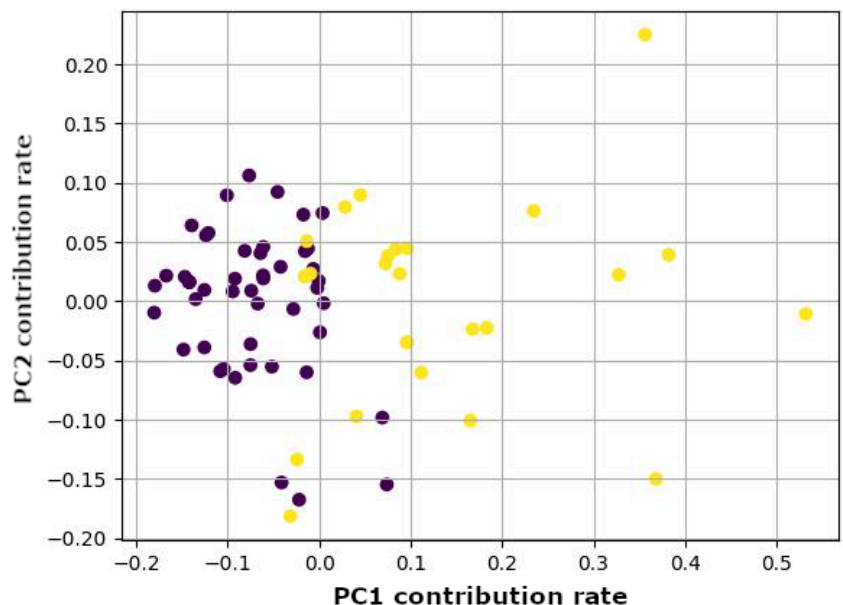

**Figure 1**. Principal Component Analysis (PCA) is the best approach to Feature Selection (dimensionality reduction); our *data are clearly separable*. As in previous results, healthy

persons (yellow) are shown to have very dispersed (variable) complexity measures, while those who are ill (dark purple) are grouped, as expected, due to the stereotypy of disease (decomplexification).

Burns and Ryan (2015) showed that it is recommendable to researchers to always use several different nonlinear measures as all of them are measuring different features of the signal. They showed that, combinations of complexity measures reveal unique information which is in addition to the information captured by other measures of complexity in EEG data'. Thus application strategy here would be to characterize the complex system (brain in this case via its output electrical signal EEG) with various nonlinear features and use statistical learning models to classify subgroups. In those applications additional biochemical, metallomics, genetic or cognitive information can improve the power of delineation between subgroups, but basic initial early detection could rely on nonlinear features based ML for screening.

**Discussion**

Our preliminary results are in line with the hypothesis that EEG-extracted HFD could be a valuable measure of early detection of an aMCI in a multimodal monitoring method aimed at early screening for high risk of conversion to an AD. We detected pathologically low complexity from resting state EEG in some participants of prior study (Eggenberger et al., 2021) that were considered to be healthy in comparison with recently published results on HFD application (Smits et al., 2016). The conclusion from the research of Smits et al. (2016) is that the low complexity measures are typically present at parietal positions, and sometimes at occipital, rarely at frontoparietal regions. We focused on the left parieto-occipital region, recommended by Smits et al. (2016) but also Babilloni (2006). If the screening methods would be based only on a couple of surface EEG electrodes, these positions would be the best candidates for detection enabling fast, cheap and accessible screening for seniors with self-reported MCI changes.

Some participants in our sample who were considered to be healthy seniors exhibited lower complexity in their EEG than the one characteristic of healthy aging according to the literature data. In fact, two of them had complexity levels lower than AD patients in the study of Smits et al, (2016). This is pointing out to the utility of this method of analysis for future early screening for aMCI as prodromal phase of AD.

Smits and her colleagues (2016) established the numerical values of EEG-extracted HFD characteristic of young adults, but also healthy aging people; in our work majority of seniors are falling in this category. Some of the participants exhibited pathologically low complexity in their signals; two persons' EEG-extracted HFD was low on all EEG positions, others showed this only on specific electrode positions (P7, P8, P4, O1, and O2), and to the lesser extent on positions P3, and Fp2, which is in line with the previous findings (Smits et al., 2016; Sao et al, 2020; Zappasodi et al,. 2014; Ahmadlou et al., 2011). Altogether, the number of participants in the study that demonstrated very low EEG complexity was below 5% of the sample, and those who showed lower complexity on some positions only, with MCI characteristic values, added to less than 20%. This early communicated information can support preventative action. This finding is also in accord with other prior research (not the complexity measures, but structural MRI) that most affected areas in prodromal phase of AD are the first in occipital lobe, and then parietal, temporal and the least in the left frontoparietal lobe (Babiloni 2007, Xia et al., 2023; Czoch et al., 2023; Sun et al,. 2020).

What we see as the most important potential impact of this work is using this detection as an early warning to neurologists who are seeing elderly patients that already report some amnestic changes (aMCI) to scrutinize those among them with pathologically low complexity by other more standard tests and find available strategies that can improve the situation gradually while majority of typically affected structures in AD are having enough neural reserve (Rossini et al, 2022). It is already shown that available cognitive training can have significant effect (Rossini et al., 2022), but dancing and exergames too (Eggenberger et al, 2020). We can confidently build on Smits and her colleagues' findings, since the methodology can be reproduced and used for further research on a larger sample in search for the most accurate early screening for aMCI as the prodromal phase of AD.

Based on our experience with previous early detection projects that rely on similar reasoning that is linking deeper compromised structures causing an elevated activity detected at cortex, the next logical step would be to combine this pathological complexity levels detection with some form of automatization or machine learning to prepare the method for everyday clinical use. Here we want to comment on numerous publications that are applying ML in their work. As we tried to demonstrate, referring to recently published work of different research groups interested in early detection of MCI and dementia in general, it is very challenging to compare those very diverse combinations of characterization of signals and various machine learning models applied, plus their different samples and sample sizers. We cannot answer straightforward whether the accuracy of traditional features based classification were higher than those nonlinear-feature-based classifications. It is essential to think of how representational they are; or, how appropriate they are for characterization of this particular signal. We can speak here about the methodological approaches that address the task differently and suggest what would improve its practical utility. For example, if classification is performed with very high accuracy (let's say above 90%) and the training set was very small, we can be pretty sure that on unseen data that accuracy would be much lover, or not informative at all. Many authors also used the mix of traditional and nonlinear measures to characterize the data, which make this comparison even more difficult. The bottom line is that we should probably think about a kind of standardization when any ML is used on biosignals to prevent overfitting and make mandatory internal and external validation, or cross-validation, and insist on much larger datasets as requirement for publication (Kohavi et al 1995; Ng, 1997; Efron&Tibshirani, 2020). What is even more recommendable is to rely on ROC use since in our opinion it is much more informative measure of model practical usefulness than accuracy. We wanted to communicate in this work that the evaluation of classifiers performance is very often missing from report of work done. What we did report here, were accuracies reported in the literature, but in our personal experience projects that relied on nonlinear measures as features tend to report lower accuracies than those that used spectral measures only, which is counter-intuitive. Very high reported accuracies like 99,7% often times underperform in real life situations, while those previously reporting 78% accuracies with more desirable EEG characterization keep having stable and useful predictions when used on unseen data. The comparison of ROC results, in our view, would be much more informative for the readers.

Half of the mentioned studies that used some EEG features to train their ML models, choose to rely on spectral measures only (Fourier based). We should bear in mind that EEG is not a stationary signal, and Fourier analysis is developed for stationary signals (it is also nonlinear and noisy which requires specific signal analysis strategy approach). Traditional spectral methods are rooted in reductionism, an analytic approach that is explaining properties of a system through properties of its elements alone, but that works only for linear systems. Not for a living systems that are demonstrating very different dynamics characterized by highly-nonlinear and nonequilibrium states (Klonowski, 2007). Despite extensive research efforts it was not confirmed that any observed changes in power spectra are correlated with changes in the level of deterministic chaos characteristic for complex systems dynamics of brain (European Parliament's report on physiological and environmental effects of non-ionizing electromagnetic radiation, from 2001).

Fractal and nonlinear measures are coming from contemporary physics methods dealing with complex systems dynamics and deterministic chaos and are increasingly important for biomedical research. They are in mathematical sense much better suited to describe nonstationary, nonlinear and noisy signals, namely electrophysiological signals. We demonstrated that nonlinear measures have excellent performance in characterizing electrophysiological signal for further machine learning classification (Čukić et al., 2020 and 2021), as features of resting state human EEG.

When we talk about fractal and nonlinear measures, there are many different families, based on various fields of research. Lempel-Ziv complexity measure, for example, is algorithmic-based measure related to Kolmogorov complexity and can be used to measure the repetitiveness of binary sequences and can establish the lossless data compression. When used to electrophysiological signal it can provide us with an information of its complexity, as regularity measure. Sample entropy (one from a big number of existing entropy measures) is determining the unpredictability or irregularity of a signal (a base for physiological interpretation that 'healthy' system is producing more irregular signals), while Higuchi fractal dimension is measuring the roughness of the signal, hence the complexity of EEG in time domain. Other measures like geometric or chaotic measures, or vast number of entropy measures are measuring different features of the same signal (applying different mathematical concepts). We cannot actually say that some of those measures are more or less important for these applications, since every one of them is extracting different kind of information from signals. Careful exploration of actual physiological mechanism (we are interested in) and prior research output can point out to most relatable methods to be applied. Burns and Ryan (2015) showed that it is recommendable to researchers to always use several different nonlinear measures as all of them are measuring different features of the signal. They showed that, combinations of complexity measures reveal unique information which is in addition to the information captured by other measures of complexity in EEG data. Thus application strategy here would be to characterize the complex system (brain in this case via its output electrical signal EEG) with various nonlinear features and use statistical learning models to classify subgroups. In those applications additional biochemical, metallomics, genetic or cognitive information can improve the power of delineation between subgroups, but basic initial early detection could rely on nonlinear features based ML for screening.

Furthermore, Berisha (et al, 2021) explained the very theoretical reasons why we always have the so-called blind spots in our data (data we never collected), and the only cure for that is – to collect more data (medical AI research projects have typically small samples, usually <100). Once deployed a highly 'accurate' model trained on a small dataset would be applied to something coming from that blind spot and fail spectacularly on classification (publicized example was BMI's Watson in cancer research). That already happened in this field, in particular classifying wrongly healthy people as MCI, and vice versa, in two consecutive classification trials with faulty model (Shafto et al., 2014.). Another preventative action is to keep monitoring the performance of the algorithm once deployed in order to be able to spot underperformance. Acosta and Topol (2022) meticulously addressed this multimodal approach to highly dimensional data collected in medicine. First, the reported accuracies are typically very high (above 80% or 90%) but it is shown that very high accuracies while training on a small dataset, usually make a lot of errors once they are tried on unseen data (the data that were not used for initial training).

The solution to these problems exists and is very thoroughly described in machine learning theory. In order to prevent unwarranted optimism, overfitting and blind spots, researchers can strategically plan certain steps to develop models that have the chance to be practically useful. First of all, the most important thing is to collect bigger samples. Since data collection in healthcare systems is typically expensive, metacentric collaborative efforts are a practical solution. Some good examples of large collaborative projects are RDoC, STAR*D, and IMAGEN (for collecting EEG data). SVM and its variants are obviously popular, but the use of embedded regularization frameworks is recommended instead (at least with the absolute shrinkage and selection operator) (Yahata et al., 2017). LOOCV and k-fold cross-validation are also popular procedures for validation (for model evaluation), and model generalization capability is typically untested on independent samples (Yahata et al., 2017). Furthermore, a Vapnik-Chevronenkis dimension (Chevronenkis, 1998) should be required as a standard for model evaluation or reduction - hence, very early when researchers make decision on methodology to be applied. Another thing is mandatory external validation in every study; without that the trained model will surely fail to classify what was planned. Without an external validation generalization of any applied ML model is questionable. Generalization is the ability of a model that was trained on one data set to predict patterns in another (unseen) data set. Testing for generalizability means that we are examining whether a classification is effective on an independent population. Yet another problem are the so-called nuisance variables, for example, those that change with age or are gender specific (Goldberger 2002; Ahmadi 2013; Ahmadlou 2013). For example, fractal dimension calculated from EEG is a little bit higher in women than in men (Ahmadi 2013). Sometimes during the development of a model, the model 'learns' some dependencies, and recognizes exactly that in unseen data, although it was not the intended task. The name of this phenomena is overfitting, and that happens when "a developed model perfectly describes the overall aspects of the training data (including all underlying relationships and associated noise), resulting in fitting error to asymptotically become zero" (Yahata 2017).

Some good examples mentioned here are using nonlinear measures for a proper characterization of EEG (Geng et al, 2022), and with a good choice of ML model (with proper internal and external validation, plus taking care of dimensionality problem), they could offer a solution for clinical recognition of prodromal phase of AD. A combination with other related factors, like proven biomarkers (for example non-ceruloplasmin copper in certain concentration, or certain cognitive tests) would be a basis for building decision support solution to augment the chances that clinicians early recognize aMCI and swiftly recommend known strategies to put in action remaining neural reserves that could increase the overall capabilities of a person.

When we presented our preliminary findings here the idea behind applying PCA was to demonstrate how clearly separable our data are, hence every ML model used would yield good enough performance on the data. Our goal to demonstrate the replicability of use of HFD as a feature (maybe together with some other entropy measures, or LZC) could lead to useful later application, was reached. We believe that future research on the topic is needed.

## Refrences